# Bichromatic microwave manipulation of the NV center nuclear spin using transition not detectable via optically detected magnetic resonance

S.M. Drofa[1,2,3], V.V. Soshenko[3,4], I.S. Cojocaru[1,3,4], S.V. Bolshedvorskii[3,4], P. G. Vilyuzhanina[1,3,5], E.A. Primak[1,2,3], A.M. Kozodaev[1,3,5], A. Chernyavskiy[1,2,3], V.G. Vins[6], V.N. Sorokin[1,2], A.N. Smolyaninov[4], S.Ya. Kilin[5,7] and A.V. Akimov[1,3,4*]

[1]Russian Quantum Center, Bolshoy Boulevard 30, building 1, Moscow, 143025, Russia
[2]Moscow Institute of Physics and Technology, 9 Institutskiy per., Dolgoprudny, Moscow Region, 141701, Russia
[3]P.N. Lebedev Institute RAS, Leninsky Prospekt 53, Moscow, 119991, Russia
[4]LLC Sensor Spin Technologies, 121205 Nobel St. 9, Moscow, Russia
[5]National Research Nuclear University "MEPhI", 31, Kashirskoe Highway, Moscow, 115409 Russia
[6]LLC Velman, 1/3 st. Zelenaya Gorka, Novosibirsk ,630060, Russia
[7]B.I. Stepanov Institute of Physics NASB, 68, Nezavisimosty Ave, Minsk, 220072 Belarus

ABSTRACT. Recently, rotation sensors utilizing the nuclear spins of nitrogen-vacancy color centers in diamond have been demonstrated. However, these devices are power-intensive and challenging to integrate into small chip-based radiofrequency antennas and circuits necessary for controlling nuclear spins or producing relatively high magnetic fields. To address this issue, the coherent manipulation of nuclear spins via coherent population trapping at moderate magnetic fields using microwave fields has been successfully demonstrated in isotopically pure diamond. In this work, we demonstrate that a similar technique can be applied to a diamond plate with a natural abundance of carbon-13, which holds significant potential for practical sensing applications. Although the forbidden resonances required for coherent control were only partially observed, coherent population trapping was successfully demonstrated at both visible and invisible transitions, with an apparent contrast of up to 98 ± 11% and a true contrast of approximately 35 ± 7%. This finding confirms the feasibility of coherent nuclear spin control even in diamond plates with naturally occurring carbon-13.

## I. INTRODUCTION

The negatively charged nitrogen-vacancy (NV) color center in diamond has garnered significant attention for its use in sensing applications. It serves as a sensitive element in various high-resolution magnetic field sensors [1–3], including scanning magnetic microscopes [4–7], fiber-based sensors [8–10], wide-field magnetic microscopes [11], and detectors capable of sensing both DC and AC magnetic fields with high sensitivity [12–15]. Additionally, the NV center can function as a local temperature sensor [16–19], a rotation sensor [20–23], a near-field fluorescence resonance energy transfer sensor [24], or even as an oriented imaging agent [25,26]. In several of these applications, the nearby nuclear spin of either the nitrogen atom or the carbon-13 isotope plays a critical role, serving as a quantum memory [16–18], an ancilla to enhance measurement fidelity [27], or as an independent sensing element [22]. However, manipulating the nuclear spin state typically requires low-frequency electromagnetic fields, on the order of one to ten MHz, which introduces the complexity of additional antennas to the experimental setup [22]. Integrating these antennas with those that generate microwave fields for electron spin manipulation [28–30] presents a challenge. These microwave fields must provide a high degree of uniformity for electron spin control while being compatible with dispersive spin-state readout methods [31]. Alternatively, a high magnetic field of around 510 G could be employed [22], but such fields are difficult to produce with minimal noise. As a result, there is a strong demand for developing methods that can control nuclear spins without relying on radiofrequency fields.

Recently, all-microwave initialization and coherent manipulation of nuclear spins have been demonstrated in isotopically pure diamond [32]. The key idea behind this approach is to apply a magnetic field that is strongly tilted relative to the NV axis, which causes mixing of state configurations and enables forbidden transitions that simultaneously flip both electron and nuclear spins. These transitions allow the manipulation of individual nuclear spins without the need for low-frequency fields. However, due to the nature of these forbidden transitions, detecting them can be challenging—even with a tilted magnetic field—since they share the same frequency as sidebands produced by hyperfine interactions with carbon-13 atoms. While the use of isotopically pure diamond can mitigate this issue, such material is expensive, making it less practical for widespread sensing applications.

The key idea of manipulation of the nuclear spin is the use of the so-called coherent population trapping, the coherent phenomena, appearing in 3-level Λ systems under biochromatic radiation [33,34]. If each of the fields, composing the biochromatic field, is in resonance with allowed transitions with a common exited state, then one could expect absorption of both fields, but instead, in the situation when the difference of two fields $\delta\omega$ is equal to the energy difference between the ground states of the $\Lambda$ – system $\Delta$, due to the quantum interference, the atomic system becomes transparent for both fields. This transmission resonance is rather narrow [35–37], limited by the width of the forbidden transition in the $\Lambda$ – systems ground state, and is usually defined by technical limitations [38–42]. In a specific case, when one of the fields is strong and another one is weak, this phenomenon is also known as electromagnetically induced transparence [43–48]. This phenomenon was successfully used for high-precision magnetometry [49], the development of compact frequency standards [50], quantum memories and photon control [51–58], and laser cooling below the recoil limit [59].

*Contact author: a.akimov@rqc.ru

In this work, we demonstrate that the coherent manipulation of nuclear spins is still achievable without directly detecting forbidden transitions when using diamond with a natural abundance of carbon. We show that despite the fact that optically detected magnetic resonance (ODMR) may not provide clear evidence of resonance, coherent population trapping (CPT) resonance can be observed with significant contrast. This is achieved by applying a biochromatic microwave field at the calculated frequency of the forbidden transition and detecting magnetic resonance. Thus, this paper paves the way for the coherent manipulation of nuclear spins in diamonds with natural carbon abundance, which are significantly more affordable than isotopically pure diamonds, by utilizing invisible transitions concealed by inhomogeneous broadening.

## II. METHODS

The ground state on an NV center is a triplet state with well-defined components of the total electron spin projection $S=1$ along the NV center axis. There is also a hyperfine interaction between the electron spin and the nuclear spin $I=1$ of nitrogen within the NV center. Magnetic transitions from the ground state with the component of an electron spin along the NV center axis $m_S=0$ to the state with $m_S=\pm 1$ alter the magnetic moment by 1. As a result, one would expect that only transitions where the nuclear spin remains unchanged are allowed $\Delta m_I=0$. However, when an external magnetic field is applied nearly perpendicular to the NV axis, the configuration of states is altered, as illustrated in Fig. 1a, enabling previously forbidden transitions [32]. For simplicity, we will refer to nuclear states by their dominant component. Transitions permitted by this dominant component are termed "allowed" and the rest "forbidden", in accordance with the basis where the quantization axis is aligned with the NV axis.

*Contact author: a.akimov@rqc.ru

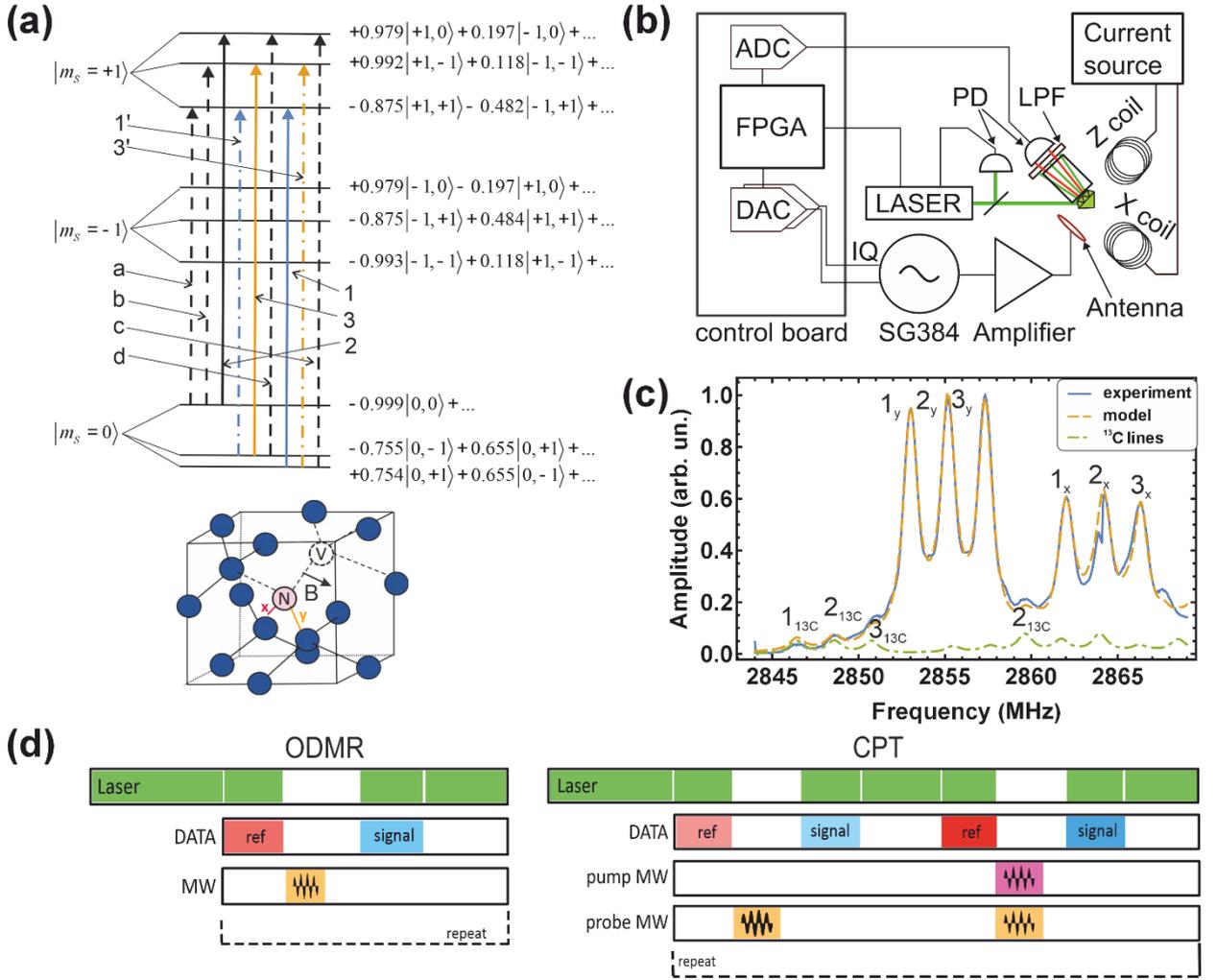

Fig. 1. (a) – Scheme of the ground state of the NV center in the external magnetic field of 30 G tilted by 88 degrees relative to the NV axis. The right column presents the configuration of states represented in eigenvectors of the basis with the quantization axis along the NV center axis, bracket notation $|m_S, m_I\rangle$ is used. Solid transition lines correspond to "allowed" transitions, dashed – to "forbidden". (b) – Scheme of experimental setup. FPGA is Spartan-6 field programmable gate array, DAC – digital-to-analog converters, IQ – quadrature modulation of microwave generator (SG384). LASER – diode laser, LPF is 650 nm longpass filter, PD – photodiode. "Z coil" and "X coil" represent two pair of Helmholtz coils….(c) – Example of the ODMR spectrum (blue solid curve) in comparison with the theoretical model, including "allowed" and isotopically shifted transitions. The orange dashed line presents the fit with the theoretical model. Magnetic field ordination is along the z-axis. The green dashed line highlights the contribution of isotopically shifted lines. (d) – Pulse sequences for ODMR and CPT experiments.

Fig. 1b illustrates the idea of the experimental setup. All measurements are performed with an ensemble of NV centers. The sample is a diamond plate with [111] cut, natural concentration of $^{13}C$, and 52 ppb concentration of nitrogen. The sample was grown by the high-pressure high-temperature method. The sample was then irradiated by an electron beam with an energy of 3 MeV and a dose of irradiation of $8 \cdot 10^{16}\,cm^{-2}$. Then it was annealed at 800 °C for 2 hours [60].

The microwave field is provided using a Helmholtz resonant antenna [60], connected to a quadrature modulator on a Stanford Research Systems SG384 generator through two amplifiers with a total gain of 63 dB. The gain of the generator is set to –32.5 dB unless otherwise specified.

Emission of the ensemble of NV centers is filtered from the excitation 520 nm light (Wavespectrum WSLD-520-001-K) with a long-pass 650 nm filter, collected by a truncated cylinder concentrator, and then is detected using a PDB-C609 Advanced Photonix photodiode with a custom amplifier.

Using the described setup, magnetic transitions in the NV center ground state are probed using pulses in the ODMR experiment, with the pulse sequence shown in Fig. 1d. The resulting spectrum is presented in Fig. 1c. For all ODMR spectra, we subtracted the fluorescence signal acquired after probing the MW pulse ("signal",

*Contact author: a.akimov@rqc.ru

Fig. 1d) from the fluorescence intensity signal, taken prior to the MW field pulse ("ref", Fig. 1d), thus suppressing the fluctuations of laser intensity and removing the offset of large NV signal fluorescence. Notably, this procedure also inverts the sign of the ODMR resonance, which typically appears as a dip in the bright fluorescence signal.

The measured ODMR spectra (e.g. Fig. 1 (c)) not only display the NV magnetic resonance triplets but also include sidebands, which arise due to hyperfine interactions with $^{13}$C nuclear spins. To account for these features, spectra are fitted using a model, where for each NV magnetic resonance triplet, a sum of nine Lorentzians is utilized. Three Lorentzians correspond to the main part of the triplet and six Lorentzians represent the sidebands. These sidebands result from interactions with two groups of $^{13}$C atoms. $^{13}$C appearing in the first group of 6 possible atomic sites near vacancy contributes to 13.7 MHz hyperfine splitting, while the second group, comprising 3 possible positions, produces a 12.8 MHz splitting [61]. To simplify the model, the average hyperfine splitting parameter of $\Delta = 13.4$ MHz was used. The formula for single orientation and the resonance model follows:

$$f(\nu) = C \left[ (1-9\alpha) \sum_{i=1}^{3} \frac{\sigma_i}{(\nu - \nu_{0i})^2 + \sigma_i^2} + \frac{9\alpha}{2} \sum_{i=1}^{3} \frac{\sigma_i}{(\nu - \nu_{0i} + \Delta/2)^2 + \sigma_i^2} + \frac{9\alpha}{2} \sum_{i=1}^{3} \frac{\sigma_i}{(\nu - \nu_{0i} - \Delta/2)^2 + \sigma_i^2} \right] . \quad (1)$$

Here $\nu_{0i}$ – the frequency of a corresponding transition, $\sigma_i$ – the width of a corresponding transition, $C$ – amplitude, $\alpha \simeq 0.012$ – the concentration of $^{13}$C, 9 stands for 9 possible positions of $^{13}$C around the NV center, and 1/2 comes from the fact that resonance contrast halves when it splits due to hyperfine interactions with $^{13}$C, which has a nuclear spin of 1/2.

In Fig. 1c, the identified peaks, associated with $^{13}$C, are marked with numbers. The control over the nuclear spin via CPT also known as electromagnetically induced transparency resonance requires a 3-level lambda system. Several such systems are possible within the NV ground state (Fig. 1a), but all of them involve forbidden transitions.

### III. RESULTS

To locate forbidden transitions in the spectra, one needs to determine the shift of energy levels in the presence of an external magnetic field, which is required to allow forbidden transitions, but also to isolate the preferred orientation of an NV center. Calculation of these shifts using a simple model with Hamiltonian:

$$H = D\hat{S}_z^2 + \gamma_e \hat{\mathbf{S}} \cdot \mathbf{B} + Q\hat{I}_z^2 - \gamma_n \hat{\mathbf{I}} \cdot \mathbf{B} + \hat{\mathbf{S}} \cdot A \cdot \hat{\mathbf{I}} \quad (2)$$

is presented in Fig. 2a and Fig. 2b. Here, $D = 2.87$ GHz and $Q = -4.945$ MHz are electron and nuclear zero-field splitting, $\gamma_e = 2.802$ MHz/G and $\gamma_n = 308$ Hz/G are electron and nuclear gyromagnetic ratios, and $A$ is the diagonal hyperfine interaction tensor with $A_{zz} = -2.162$ MHz and $A_{xx} = A_{yy} = -2.62$ MHz [62]. The operators $\hat{\mathbf{S}}$ and $\hat{\mathbf{I}}$ are vector operators for electron and nuclear spins, respectively. It can be observed that allowed transitions with isotope shift almost always overlap forbidden transitions at small tilt angles (see Fig. 2a). For near-perpendicular tilt angles, though, they do not overlap as long as allowed transitions experience minor splitting. However, the strength of all transitions decreases at a 90° angle, so balance should be maintained. Small forbidden transition probability leads to higher requirements for microwave power to observe the same ODMR contrast or smaller ODMR contrast if the power is maintained the same. Thus, to detect $c$ transition, the external field for the observation of forbidden transitions on ODMR is tilted at about 89–89.9° relative to the NV-center axis, where lines, corresponding at 0 magnetic field to c transition and "3, $^{13}$C" (which is transition "3" for level $m_s = +1$ shifted by the presence of $^{13}$C nearby), are splitting (see Fig. 2b).

*Contact author: a.akimov@rqc.ru

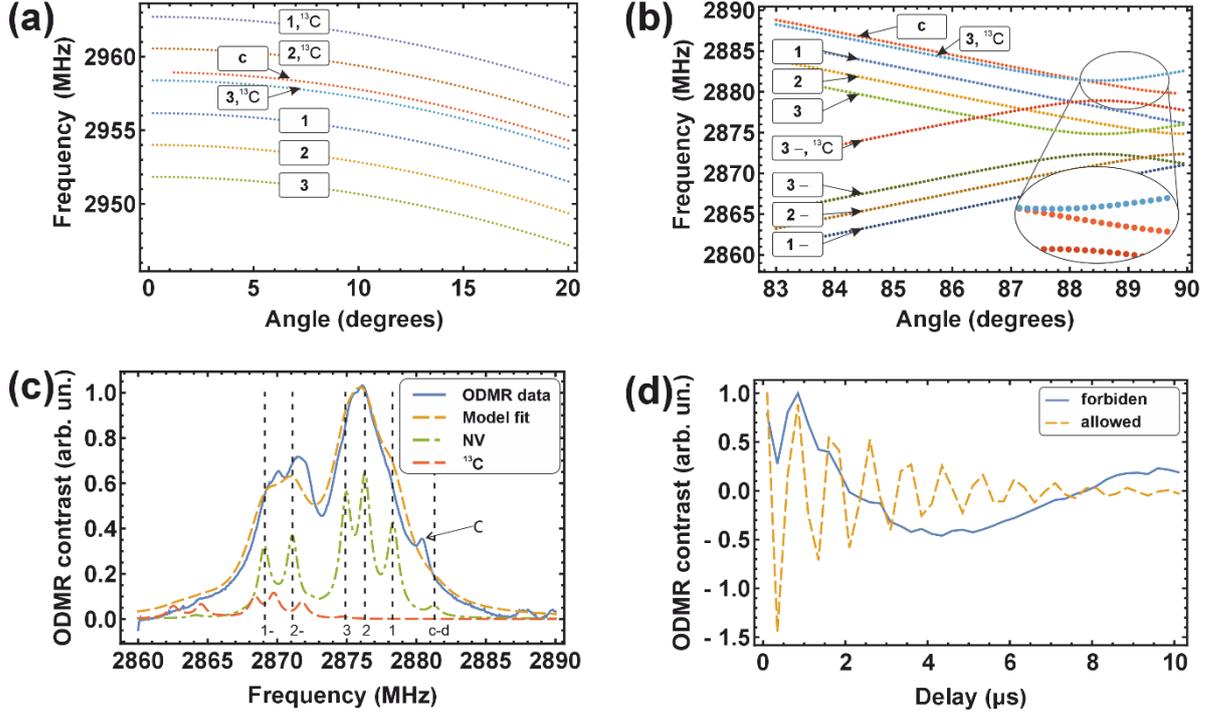

Fig. 2. (a) Model of a Zeeman shift of the NV center oriented along 111 axis versus the angle of the total applied magnetic field, relative to the NV center. (b) Model of a Zeeman shift of 111 oriented NV center levels versus the angle of the total applied magnetic field, relative to the NV center. Transitions "1-", "2-", "3-" are associated with $m_S = 0 \leftrightarrow m_S = -1$ transitions, where the nuclear substate is the same as in transitions "1", "2", "3". (c) Comparison between ODMR (blue) perpendicular to the NV-center main axis field with a 2-level model with no isotopic shift (orange dashed). The contribution of $^{13}C$ within the two-level model is presented in a green dot-dashed line, while forbidden transitions (degenerate c, d) are marked in red long and short dashed lines. The amplitude of the ODMR peaks was normalized to unity. (d) Rabi oscillations for the forbidden transition c (Blue) with a generator set to -30 dBm. For comparison, there are Rabi oscillations for strong transition 3 (orange dashed line). The scale of the allowed amplitude of transition oscillations to forbidden transition oscillations is 10 to 1, while oscillations were observed under the same microwave conditions. The amplitude of the signal is normalized to unity with the equilibrium signal level being taken as 0.

Once the position of the transitions in the given field is identified, one can compare calculations with the ODMR spectrum (see Fig. 2c, Table 1). Only one of the forbidden resonances could be detected, namely the "c-d" resonance (see Fig. 1a). This transition could be driven with a microwave field, as shown in Fig. 2d. It is evident that the transition is driven considerably slower than the allowed transition "3", and shows a trace of off-resonance Rabi oscillations from nearby allowed transitions due to significant microwave power.

| Transition | Frequency, MHz |
|---|---|
| 1- | 2869.1 |
| 2- | 2871.1 |
| 3 | 2874.9 |
| 2 | 2876.3 |
| 1 | 2878.3 |
| c-d | 2881.3 |

Table 1. Transition frequencies of the NV center in a near-perpendicular magnetic field

We note here that the angle of the magnetic field of 89.5° with respect to the 111 axis was used to detect the transition, but it is not ideal for further experiments, since at this angle, ODMR triplet resonances are not fully resolved, and populations of different hyperfine levels are mixed. To maintain distinct triplet resonances and adhere to the traditional ODMR framework, while still allowing the forbidden transitions to remain weakly allowed, we reduced the angle of the magnetic field to 87.6° with respect to the 111 axis in subsequent experiments.

Transition "c" can form a 3-level "V" system with transition "1", as shown in Fig. 3a, and a 3-level "Λ" system with transition "2", while transitions "a" and "b" form a similar 3-level "Λ" system with transitions "1" and "3", respectively.

*Contact author: a.akimov@rqc.ru

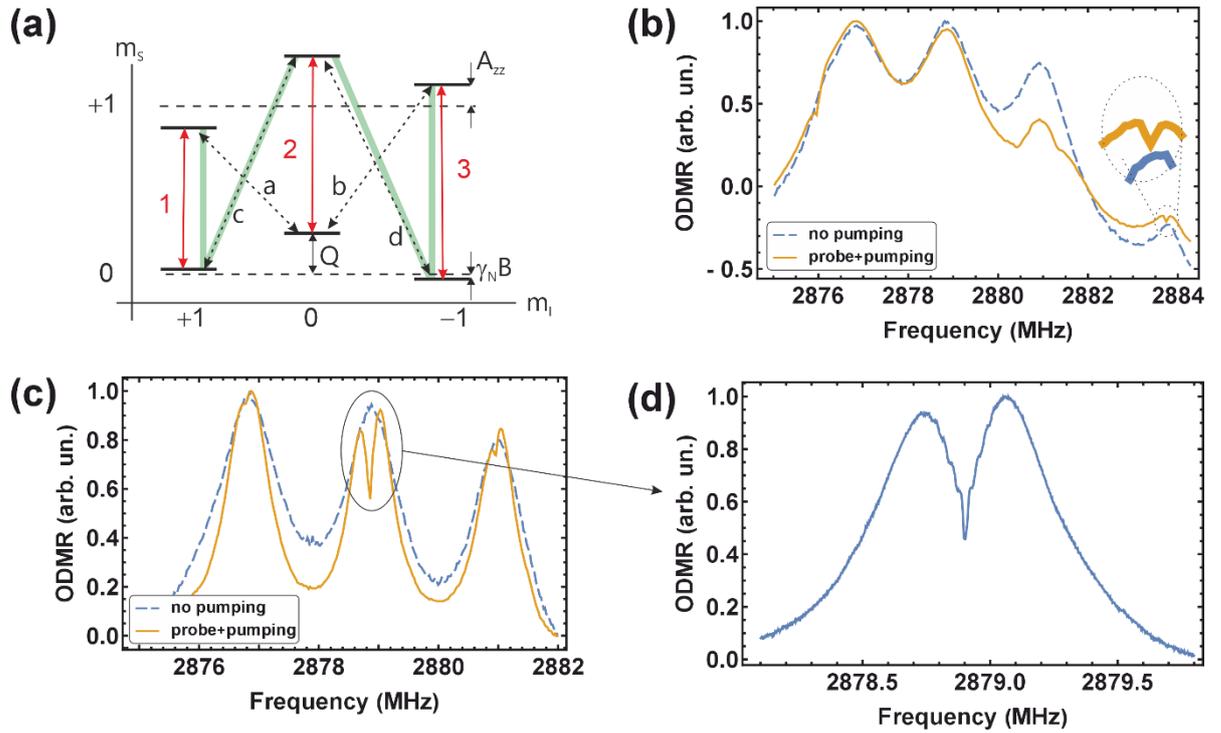

Fig. 3. (a) – Ground state of the NV center in a "perpendicular" magnetic field. Thick color lines indicate the V-systems associated with "c", "d" transitions. (b) CPT resonance for pump pulse set to "3" transition, the probe pulse was scanned in frequency. (c) ODMR in comparison to CPT. – Pump pulse is set to forbidden transition "c", the probe is scanned in frequency. (d) The detailed spectrum of transition 2 in the presence of radiation at transition "c".

To observe the CPT signal, one of the microwave fields, specifically the pump pulse, was set to transition "3", and the second microwave field, probe pulse, was scanned in frequency around the frequency of nearly degenerate transitions "d" and "c". The fluorescence signal was detected during the probe pulse, and the collection time was set to 130 μs to maximize the signal-to-noise ratio. As one can see in Fig. 3b, there is significant suppression of one of the nitrogen nuclear spin related resonances with the nuclear spin $m_I = -1$. This is likely caused by partial optical pumping from the states $m_I = -1$, $m_I = 0$ to $m_I = +1$ by microwave radiation permanently tuned to transition "3". While transition "2" is not in direct resonance with MW radiation, its wing is still excited, and therefore we expect that over a long time, the population will be mostly collected on $m_I = +1$ state. Moreover, on top of transition "d" itself, one can see a weak narrow dip in the ODMR signal. If one instead uses a pump pulse centered at transition "d" and observes transition "3" with a probe pulse, then the CPT resonance is much easier to see on top of the strong transitions "2" and "3" (see Fig. 3c). The strongest of the resonances, corresponding to the transition "2", is demonstrated in Fig. 3d in more detail. The resonance clearly has some modulation of its slopes, which will be discussed below. The dip at resonance "2" cannot be explained by the "V" scheme, presented in Fig. 3a, and corresponds to the Λ scheme, presented in Fig. 4a. The Λ scheme indeed tends to have a much bigger contrast most likely due to the fact that $m_S = -1$ as a result of interaction with green light (see Fig. 1d) has a significant chance to decay at $m_S = 0$ thus losing nuclear spin coherence, while $m_S = 0$ only couples to the excited state and thus its coherence is less disturbed.

While only one transition is visible in the spectrum, one can identify the positions of the rest using the calculations, presented in Fig. 2a. If one sets one of the microwave fields to the calculated frequency of transition "a" and the other on the frequency of strong transition "3", then the Λ resonance with invisible transition must be observed. The precise location of the transition may be detected by the CPT resonance itself. To reveal the position of the CPT peak, we use a 2D scan plot with both "pump" and "probe" microwave radiation frequencies scanned (Fig. 4b) and thus allow the precise location of the resonance.

As shown in Fig. 4c, the resonance is almost as strong as for transition "d" despite the fact that transition was not detected in the ODMR spectrum. It is interesting to note that for the case of resonance related to the "d" transition, both Λ schemes of transition "2" in pair with "c" and in pair with "d" are observed at the same frequency. Thus, contrasts of both "a"-related and "d"-related CPT resonances are comparable.

One can ask how CPT resonances have more or less the same contrast, despite transitions themselves being so weak that they cannot be directly observed. The explanation can be provided as follows: it is true that for both "a" and "d" resonances, the microwave intensity and dipole moment of transitions are similar and therefore have similar contrast Rabi frequencies of transitions "a" and "d". Here, CPT resonances are assumed to be far from saturation. Nevertheless, it is not necessary to contradict the fact that the resonance was not observed. The fact that transitions "a" and "b" are not observed at the ODMR spectrum may be explained because they are very close to the strong transitions 2 and 3, both with the isotopic shift of 13C in the nearest node, which is due to their relatively large width in natural diamond.

*Contact author: a.akimov@rqc.ru

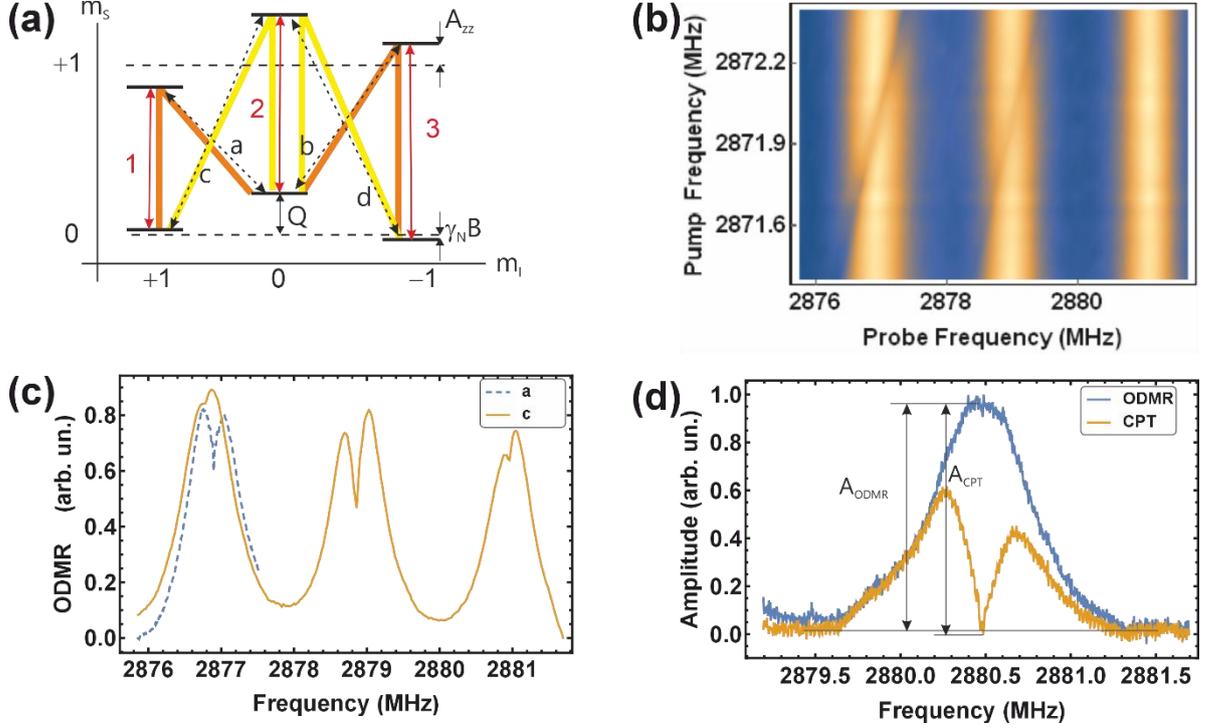

Fig. 4. (a) – Ground state of the NV center in a "perpendicular" magnetic field. Thick color lines indicate the $\Lambda$-systems associated with "c" and "d" transitions. $Q = -4.945$ stands for nuclear zero-field splitting, $A_{zz} = -2.162$ MHz is the longitudinal component of the hyperfine interaction tensor. (b) Two-dimensional scan of ODMR in the presence of two microwave fields. (c) Comparison of CPT resonances with pump pulse set to "a" (blue) and "c" (orange) transitions. (d) – Comparison of ODMR resonance and ODMR resonance with the CPT dip associated with "c" transition.

The contrast of CPT resonance could be improved if larger microwave radiation power is used. Fig. 5a illustrates the rise of the contrast with the power of microwave radiation when both microwave fields were rising at the same time. We should note that the contrast in the case of CPT resonance is loosely defined here. The naïve definition of "apparent contrast" $C_{ap}$ would be just to compare the amplitude $A_{ODMR}(f_r)$ of ODMR resonance with one MW field to the level of the CPT resonance minimum with respect to non-disturbed ODMR resonance $A_{EIT(f_r)}$ as indicated in Fig. 4d. Thus, the apparent contrast is defined as:

$$C_{ap} = \frac{A_{EIT(f_r)}}{A_{ODMR}(f_r)} \quad (3)$$

Nevertheless, it may also be useful to compare the low level of ODMR resonance to the level without MW at all. As demonstrated in Fig. 5a, the formation of ODMR resonance is a dip in fluorescence, which is induced by the transfer of part of the population from $|m_s = 0\rangle$ state to $|m_s = 1\rangle$ state. Thus, if NV does not absorb MW radiation at all, like it should if all the population is in the CPT state, then the emission of the NV center should be maximal and equal to the NV emission without microwave radiation. However, due to the fact that ODMR resonance is always a dip in the emission, if we add one "pump" microwave field, the emission of NV should drop already, thus creating a new base level for ODMR resonance. On top of this, in the ensemble of NV centers, there should be many non-homogenous contributions to the ODMR and CPT resonances, respectively, but the position of CPT resonances as long should be the same as the ground state has no inhomogeneous broadening [63,64]. Thus, the CPT dip should add up from all inhomogeneous contributions and in principle go to the level of NV emission with no MW. In this sense, the apparent contrast should actually be over 100%. The true contrast should be defined as (see Fig. 5b):

*Contact author: a.akimov@rqc.ru

$$C = \frac{A_{EIT(f_r)}}{A_{ODMR}(f_r) + \Delta A} \qquad (4)$$

and is considerably smaller than the apparent contrast in our experiments. In this formula, $f_r$ is the frequency of the CPT fringe, $A_{EIT}(f)$ is the profile of ODMR under pump and probe pulses, and $A_{ODMR}(f)$ is the profile of classic ODMR. Here we used max $A_{EIT}$ for contrast estimation, because precise measurement of ODMR is obstructed by background change due to probe pulse and pulse length being not equal to π-pulse.

There are two important features of the CPT resonance at large microwave power: first, it is apparent that resonance has some structure, which on one side has splitting at the center of the resonance, on another periodic modulation of the resonance wings (Fig. 6). The period of modulation of the resonance wings does depend on the magnetic field but does not depend on microwave power. It is interesting that the frequency of these oscillations is very close to the frequency of Larmor precession of the carbon-13 nuclei surrounding the NV center, which for the magnetic field $B$ is [65].

$$\omega = \frac{g_{C13}\mu_N B}{\hbar} = 2\pi \cdot (535 \text{ Hz/G}) B . \qquad (5)$$

According to this formula, the frequency for a magnetic field of 30 G is 32 kHz and for 45 G is 48 kHz, which is close to the experimentally found values: 30 kHz and 45 kHz, respectively. The physical picture behind this correspondence remains to be discovered. Resonance modulation for bichromatic NV center excitation was previously studied in the work [66].

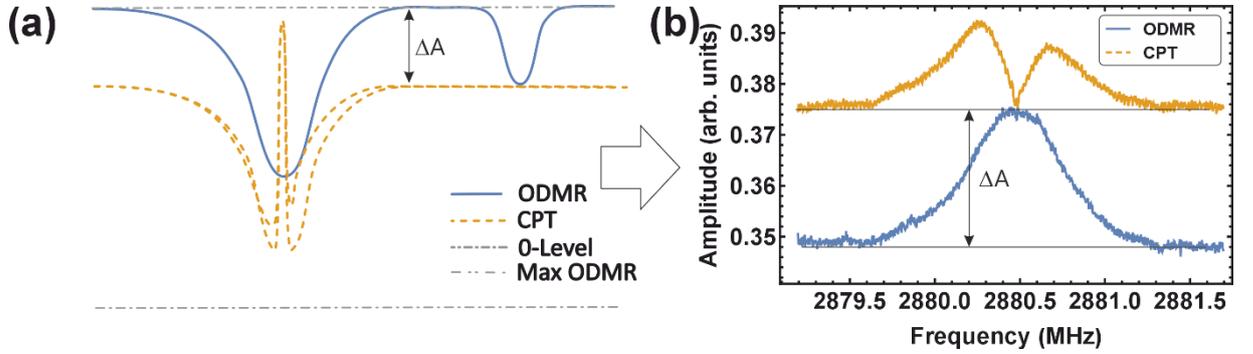

Fig. 5. (a) – Schematic illustration of the formation of ODMR resonance under 1 (blue) and 2 (orange) MW fields. Orange lines also illustrate inhomogeneous broadening. (b) – Same as Fig. 4d but with the use of only the first reference (see Fig. 1d right) for both CPT and single MW ODMR resonance.

The splitting of the resonance center was not carefully investigated due to its small value compared to the other features. It is most likely caused by the excitation of both "c" and "d" transitions, as the applied microwave field to "c" and "d" transitions corresponded to a Rabi frequency of about 60 kHz while splitting between levels $|0,+1\rangle$ and $|0,-1\rangle$ at 45 G is an order of 10 kHz.

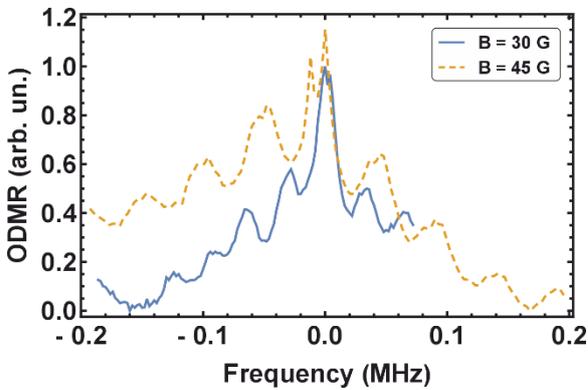

Fig. 6. Zoom of CPT resonance associated with the "a" transition for two magnetic fields: 30 G (blue) and 45 G (orange).

After the optimization of the MW pulse length, microwave power, and green laser power, as well as the optimization of the ratio of amplitudes of the pump to probe pulses, an apparent contrast of 98±11% for resonance associated with transitions "c" and "d" was achieved (see Fig. 4d). Nevertheless, the true contrast in this case is only 35 ±7%.

### IV. CONCLUSION

A diamond sample with a natural abundance of carbon-13 and 0.93 ppm of nitrogen was investigated under bichromatic microwave radiation in the presence of laser pumping. In magnetic fields, nearly perpendicular to the axes of NV centers selected for investigation, CPT resonances on V and Λ systems involving forbidden transitions were observed despite the fact that forbidden resonances were not observed in the ODMR resonances. The power dependence of resonances contrast was investigated and found to be linear within available MW powers. The resonances were found to be modulated by amplitude. The modulation frequency depends on the magnetic field and was in good correspondence with the precession frequency of the carbon-13 spins. Furthermore, the contrast of CPT transition was optimized reaching an apparent contrast of 98 ± 11% and a true contrast of 35 ± 7%, paving the way for purely microwave field control of nuclear spins in NV centers and radio-frequency free rotation sensors.

*Contact author: a.akimov@rqc.ru

## ACKNOWLEDGMENTS
This research was supported by a grant from the Ministry of Science and Higher Education of the Russian Federation No. 075-15-2024-556.

*Contact author: a.akimov@rqc.ru

*Contact author: a.akimov@rqc.ru

*Contact author: a.akimov@rqc.ru